\documentstyle[12pt]{article}
\input{epsf}
\def\la{\;\raise0.3ex\hbox{$<$\kern-0.75em\raise-1.1ex\hbox{$\sim$}}\;}
\def\ga{\;\raise0.3ex\hbox{$>$\kern-0.75em\raise-1.1ex\hbox{$\sim$}}\;}
\newcommand{\tn}{\mbox{$T_{cn}$}}
\newcommand{\tp}{\mbox{$T_{cp}$}}
\newcommand{\te}{\mbox{$T_{\rm e}$}}

\begin{document}

\title{\bf Standard and Enhanced Cooling of Neutron Stars with Superfluid Cores}

\author{K. P. Levenfish and D. G. Yakovlev \\
    {\it   Ioffe Physical Technical Institute} ,\\
    {\it Russian Academy of Sciences,} \\
    {\it Politekhnicheskaya 26, St. Petersburg, 194021 Russia} }

\date{June 13, 1995}
\maketitle

\begin{center}
(Published in {\it Astron. Lett.}, v. 22, No. 1, 56--65, 1996)
\end{center}

\begin{abstract}
 Calculations are performed of
the cooling of neutron stars with standard and enhanced neutrino
energy losses in the presence of neutron and
proton superfluidities in the stellar cores.
The effects of superfluidity on the heat capacity and the
neutrino luminosity produced by the direct and
modified Urca processes and by the neutrino bremsstrahlung emission in
nucleon-nucleon collisions are taken into account.
The constraints are analyzed
on the critical temperatures $\tn$ and $\tp$
of the transition of neutrons and
protons to the superfluid state, which can be obtained
from a comparison of observational data on thermal radiation from neutron
stars with theoretical cooling curves for the standard and
enhanced neutrino energy losses.
  Possible ranges of $\tn$ and $\tp$ for the pulsar Geminga are discussed.
\end{abstract}

\section{INTRODUCTION}

Neutron stars (NSs) are unique astrophysical objects.
The density in the NS
central layers is several times higher than the  standard nuclear
density $\rho_0 = 2.8 \times 10^{14}$~g$\cdot$cm$^{-3}$. Such a dense matter
cannot be analyzed in laboratories, but it
can be studied by astrophysical methods, in particular,
by comparing theoretical
calculations of the cooling of NSs with observational data on their thermal
emission.

The nuclear
composition of the NS central layers is not known with certainty
(Shapiro and Teukolsky 1983). We will restrict our consideration
to the simplest model in which  matter
is composed of neutrons ($n$), protons ($p$), and
electrons ($e$). The cooling rate of a not-too-old star
(with an age $t < 10^5$~yr) is determined by the neutrino luminosity of
its inner layers.
One can distinguish two types of NSs, with regard to the efficiency
of the neutrino energy losses.
Stars with a relatively low ({\it standard}) energy loss rate (see, e.g.,
Friman and Maxwell 1979; Yakovlev and Levenfish 1995) caused by the modified
Urca processes
\begin{equation}
    n + N \rightarrow p+N+e+ \bar{\nu}_e, \; \;
    p + e+ N \rightarrow n + N + \nu_e,
\label{eq:Murca}
\end{equation}
and by the bremsstrahlung radiation of neutrino pairs in nucleon-nucleon
($N$) collisions
\begin{equation}
    N + N \rightarrow N + N + \nu + \bar{\nu}.
\label{eq:Brema}
\end{equation}
constitute the first type of NS.

A buffer nucleon $N$ in reaction (\ref{eq:Murca}) is needed
to satisfy momentum conservation in dense matter with relatively low
number densities of $e$ and $p$.
To avoid misunderstandings, note that we are dealing here with highly
degenerate neutrons and protons.
The neutron branch of the reaction ($N = n$) is
well known; the proton branch ($N = p$) has a comparable intensity, which has
been pointed out only recently (Yakovlev and Levenfish 1995).
Rather massive NSs whose neutrino luminosity is enhanced by
the direct Urca process (Lattimer {\it et al.} 1991)
\begin{equation}
    n  \rightarrow p+e+ \bar{\nu}_e, \; \;
    p + e \rightarrow n + \nu_e.
\label{eq:Durca}
\end{equation}
belong to the second type.

At a not too high density
$\rho \la \rho_0$, this process is forbidden by
momentum conservation.

Observational data on thermal emission from at least three pulsars --
PSR$\,0656+14$ (Finley {\it et al.} 1992; Anderson {\it et al.}, 1993),
PSR$\, 0833-45$
(\"{O}gelman and Zimmermann 1989), and Geminga (Halpern and Holt 1992;
Halpern and
Ruderman 1993; Meyer {\it et al.} 1994) --
indicate that their cooling is faster
than the {\it standard} cooling but slower
than that enhanced by the process (\ref{eq:Durca}).
In addition, the cooling can be controlled by possible
superfluidity of neutrons and/or protons in the stellar core.
Superfluidity with a moderate critical temperature
$T_c$ ($0 < (T_c$--$T) \la T_c$,
where $T$ is the temperature of the matter)
suppresses the neutrino energy losses
but increases the heat capacity through
the release of latent heat in a phase
transition. Strong superfluidity $(T_c \gg T)$
exponentially suppresses both the
neutrino losses and the heat capacity of the NS.
It is the interplay of these
factors that may speed up the standard cooling or
slow down the enhanced cooling.

The main parameters of superfluidity that
affect the cooling are the critical
temperatures $\tn$ and $\tp$ of the
transition of neutrons and protons, respectively, to a superfluid state.
A rigorous calculation of $\tn$ and $\tp$ in the
core of a NS requires a consistent many-particle
quantum theory that adequately
describes strong particle interactions. Such a theory has not yet been
constructed, but a large number of theoretical
models of superdense matter have
been proposed. The temperatures $\tn$ and $\tp$
calculated in various models (see
Wambach {\it et al.} 1991 and Page 1994 and references therein) vary
noticeably with density in the range $10^7$--$10^{10}$~K.
By comparing observations of thermal
emission from isolated NSs with calculations of their cooling at various
temperatures $\tn$ and $\tp$, one can place independent constraints on the
parameters of neutron and proton superfluidity in the NS core.

A number of studies are devoted to the models of
the standard cooling of a NS with a
superfluid core (see, e.g., Nomoto and Tsuruta 1987 and references therein).
Page and Applegate (1992) were the first
to calculate the cooling enhanced by
the direct Urca process in neutron stars with a superfluid core. In these
studies, the superfluidity of one nucleon species was
considered, and rather approximate fitting
formulas describing the effect of superfluidity on the heat capacity and
neutrino luminosity were used.
A consistent analysis of the effect of $n$ and $p$
superfluidity on the heat capacity and
neutrino energy losses in the
reaction (\ref{eq:Durca})
was carried out by Levenfish and Yakovlev (1993). Gnedin and Yakovlev (1993)
used these results and, following Page and Applegate (1992),
simulated numerically the enhanced cooling of a NS in the presence
of neutron or proton superfluidity. The
enhanced cooling with both $n$ and $p$ superfluidities at once was
first considered by Van
Riper and Lattimer (1993), who based their analysis on very
approximate factors
of suppression of the neutrino luminosity by superfluidity. A consistent
calculation of the above factors was performed by
Levenfish and Yakovlev (1994a,b).
Gnedin {\it et al.} (1994) used these results to model the enhanced cooling of
the pulsar PSR$\, 0656+14$ in the presence of a combined nucleon
superfluidity in
its core. Comparison of these calculations with observational data on the
pulsar's thermal emission allowed the authors
to obtain constraints on the critical
temperatures $\tn$ and $\tp$ in the pulsar's core (assuming that the neutrino
energy losses are enhanced by the direct Urca process).

In this paper, we continue to study the cooling of neutron stars with a
simultaneous superfluidity of nucleons.
Apart from the enhanced cooling, we also consider
the standard cooling. The suppression factors for neutrino energy
losses required for this consideration have
recently been obtained by Yakovlev
and Levenfish (1995).

Our models are applied to the pulsar 2CG$\, 195+04$ (Geminga). The effects of
neutron and proton superfluidities on the standard or enhanced cooling of
Geminga were first investigated by Page (1994), who used approximate
suppression factors for the heat capacity
and neutrino energy losses. He showed
that observations could be explained for any type of cooling
by assuming that the stellar core was superfluid. We
develop more detailed models of the NS
cooling by including new data on the neutrino energy
losses due to the proton branch of the
process (\ref{eq:Murca}) (Levenfish and Yakovlev 1993, 1994a,b;
Yakovlev and Levenfish 1995).

\section{MODELS OF COOLING NEUTRON STARS}

We considered two models of a NS.
In both cases, the equation of state of matter
($npe$) in the stellar core was taken from Prakash {\it et al.} (1988).
The maximum
mass of the NS for this equation of state is $1.7\; M_\odot$.
In the first model, the
mass of the star is $M = 1.44\; M_\odot$, the radius is 11.35 km,
and the central density is $\rho_c = 1.37 \times 10^{15}$~g$\cdot$cm$^{-3}$.
In the second model, $M = 1.3\; M_\odot$, ${\cal R} = 11.71$~km,
and $\rho_c = 1.12 \times 10^{15}$~g$\cdot$cm$^{-3}$. The chosen equation of
state allows the direct Urca process to operate at
densities $\rho > \rho_{cr} = 1.30 \times
10^{15}$~g$\cdot$cm$^{-3}$. Thus, the first model corresponds
to the enhanced cooling: the
powerful direct Urca process is possible in a
small central kernel with the radius of
2.32~km and mass of 0.035~$M_\odot$. In the second model, the critical
density $\rho_{cr}$ is not achieved, and
the star has the standard neutrino luminosity.

We assumed that nucleons could be superfluid
throughout the entire NS core. The
proton superfluidity results from Cooper pairing of protons in the
$^1S_0$ state, while the neutron superfluidity is due
to the neutron pairing in the $^3P_2$ state with zero
projection of the pair's moment onto the quantization axis (see, e.g.,
Wambach {\it et al.} 1991).
The critical temperatures $\tn$ and $\tp$ can depend on
density of matter and are not definitely known
(Section 1). For certainty,
we assumed $\tn$ and $\tp$ to be constant throughout the stellar core.

For our cooling calculations, we used the program (Gnedin and Yakovlev 1993;
Gnedin {\it et al.} 1994) in which the inner layers of the NS were assumed to be
isothermal. This approximation is valid
for a not-too-young NS ($t > 10$--$10^3$~yr)
whose internal thermal relaxation has already occurred.
Following Glen and Sutherland (1980),
the isothermal region was specified by
the condition $\rho > \rho_b = 10^{10}$~g$\cdot$cm$^{-3}$.

The photon luminosity of the
NS depends on the
effective temperature $\te$ of the
stellar surface. The relation between $\te$ and the temperature $T_i$ near
the
boundary of the isothermal region ($\rho = \rho_b$) is determined by the
thermal insulation of the outer
envelope ($\rho < \rho_b$). The magnetic field
in the envelope causes the heat conduction to be anisotropic,
resulting in an anisotropic distribution of
$\te$ over the NS surface (see, e.g.,
Yakovlev and Kaminker 1994). Even a relatively weak field $B > 10^{10}$~G
magnetizes degenerate electrons and reduces
the heat conduction across {\bf B}. This
enhances the heat insulation of the NS
in the vicinity of the magnetic equator.
A stronger field, $B > 10^{12}$~G, is quantizing
both for a degenerate electron gas in deep layers of the envelope
and for nondegenerate sub-photospheric layers. The
effects of quantization of electron orbits
generally increase the longitudinal
thermal conductivity
of degenerate electrons, reduce the pressure of
degenerate electrons
(in the limit of strong quantization), and enhance the
radiative thermal conductivity
of nondegenerate layers. As a result, the quantizing field
weakens the heat insulation near the magnetic poles.
The NS cooling has been commonly
calculated under the assumption that the magnetic field
is radial in every point
of the surface (see, e.g., Van Riper 1991 and references therein). In this
case, only the reduction of the thermal insulation by the quantizing magnetic
field was taken into account.
Recently, Shibanov and Yakovlev (1996) have
considered the cooling of a NS with a
dipole field by taking into account both
the reduction of the thermal insulation near
the magnetic poles and the enhancement of the insulation near the equator.
Both effects almost compensate each other for a field strength at
the magnetic pole equal to $B \approx 3 \times 10^{12}$~G, and the cooling
proceeds in the same way as for $B = 0$, although the magnetic field
gives rise to a strong anisotropy in $\te$. Our calculations
are applied to the Geminga pulsar
(Section~1) whose magnetic field is close
to the above value (see, e.g., Bertsch {\it et al.\ } 1992). Therefore, we
neglect the effect of the magnetic field and
use the relation between $\te$ and $T_i$ derived by Van Riper (1991)
for $B = 0$. In this approximation,
by $\te$ we mean the average effective surface temperature
which determines the total photon luminosity of the star
$L = 4 \pi {\cal R}^2 \sigma \te^4$ (without allowing for the gravitational
redshift). Note that in the calculations of
Gnedin {\it et al.} (1994), like in other calculations
in which the magnetic field has been
considered to be radial, the influence of
the field on the cooling is highly overestimated.

The neutrino luminosity of the
NS with $M = 1.44\, M_\odot$ is mainly determined by the
direct Urca process (\ref{eq:Durca}). The rate of the corresponding
neutrino energy losses in
normal (nonsuperfluid) matter was
calculated by Lattimer {\it et al.\ } (1991).
Levenfish and Yakovlev (1993)  calculated
the factors $R$ of suppression of the reaction (\ref{eq:Durca})
by the neutron superfluidity for normal protons ($R = R_n^{\rm (d)}$) and
by the
proton superfluidity for normal neutrons ($R = R_p^{\rm (d)}$).
They also calculated (Levenfish and
Yakovlev, 1994b) the suppression of the
reaction (\ref{eq:Durca}) by both superfluidities ($R =
R_{np}^{\rm(d)}$). It turned out that
the reaction is mainly suppressed by
the strongest superfluidity,
$R_{np}^{\rm (d)} \sim \min[R_n^{\rm (d)},\; R_p^{\rm (d)}]$.
Note that Van Riper and Lattimer (1993)
used the
approximate suppression factors $R_{np}^{\rm (d)} \sim R_n^{\rm (d)}
\cdot R_p^{\rm (d)} $, which are
especially inaccurate if $R_n^{\rm (d)} \sim R_p^{\rm (d)} \ll 1 $.
In addition to the powerful direct Urca process, we also took into
account the standard neutrino reactions (\ref{eq:Murca}) and
(\ref{eq:Brema}).

The neutrino luminosity of
the $M = 1.3\,  M_\odot$ star
is determined by both the
modified Urca processes (\ref{eq:Murca}) and
the neutrino emission in nucleon collisions (\ref{eq:Brema}) in
the entire stellar core. Friman and Maxwell (1979) considered the
neutron branch of the modified Urca process
(the process (\ref{eq:Murca}) for $N = n$) and the
bremsstrahlung emission of nucleons (\ref{eq:Brema})
in nonsuperfluid matter.
Yakovlev and Levenfish (1995) studied the proton branch of the modified Urca
process ($N = p$) in nonsuperfluid matter
and showed that this branch
was almost as efficient as the neutron branch.
Note that the proton branch was
commonly considered to be weak and
excluded from calculations of the NS cooling.
Yakovlev and Levenfish (1995) also calculated
the suppression factors of the proton
($R = R^{({\rm m}p)}$) and neutron ($R = R^{({\rm m}n)}$) branches of
the reaction (\ref{eq:Murca})
by the neutron
superfluidity ($R^{({\rm m}n)}_n$ and $R^{({\rm m}p)}_n$)
for normal protons and
by the proton superfluidity ($R^{({\rm m}n)}_p$ and $R^{({\rm m}p)}_p$)
for normal neutrons.
The suppression factors for the reactions with the same
number of superfluid particles turned out to be rather close if they were
compared as functions of the corresponding dimensionless
energy gap $v= \Delta_N/T$.
Here, $\Delta_p$ is the energy gap for the
proton (singlet) superfluidity, and
$\Delta_n$  is the minimum energy gap at the Fermi surface of neutrons
for the anisotropic (triplet) neutron superfluidity.
Thus, for example, the factors $R_p^{\rm (d)}(v_p)$
and $R_p^{({\rm m}n)}(v_p)$ are close to each other
and to the factors
$R_n^{\rm (d)}(v_n)$
and $R_n^{({\rm m}p)}(v_n)$.
Taking this fact into account, we propose
approximate factors of
suppression of the neutron ($R=R_{np}^{({\rm m}n)}$)
and proton ($R=R_{np}^{({\rm m}p)}$) branches of the
modified Urca process by a combined nucleon superfluidity. One can
expect these factors as functions of the
corresponding $v$ (corrected for
the number of superfluid particles involved)
to be not
too different from the factor $R^{\rm (d)}_{np}$
for the direct Urca process. Therefore, we use
the approximate expressions
$R^{({\rm m}p)}_{np} (v_p,v_n) \approx R^{\rm (d)}_{np}(2v_p,v_n) \,
R^{({\rm m}p)}_n(v_n) / \,R^{\rm (d)}_n(v_n)$ and
$R^{({\rm m}n)}_{np} (v_p,v_n) \approx R^{\rm (d)}_{np}(v_p,2v_n) \,
R^{({\rm m}n)}_p(v_p) / \, R^{\rm (d)}_p(v_p)$.
If the protons are normal
($v_p=0$), the expression for $R^{({\rm m}p)}_{np} (v_p,v_n)$
becomes exact, while for normal neutrons
($v_n=0$) the factor $R^{({\rm m}n)}_{np} (v_p,v_n)$ becomes exact.

The neutrino energy loss rates
in the nucleon-scattering processes (\ref{eq:Brema})
in nonsuperfluid matter are
taken from Friman and Maxwell (1979).
The loss rate in the $pp$-scattering is calculated
from equation~(52) of this paper  for the
$nn$-scattering with the neutron parameters replaced by the proton parameters
(see Yakovlev and Levenfish 1995, for details).
The factors of suppression of the
$pp$- ($R = R^{(pp)}_p$) and $np$-scattering ($R = R^{(np)}_p$)
by the proton superfluidity were calculated by Yakovlev and Levenfish (1995).
The similarity relations mentioned above
allow us to propose the approximate suppression factors $R^{(nn)}_n$
for the $nn$-scattering by the
neutron superfluidity and
the suppression factors $R^{(np)}_{np}$ for the $np$-scattering by
the combined nucleon superfluidity:
$R^{(nn)}_n \approx R^{(pp)}_p(v_n)$ and
$R^{(np)}_{np} \approx R^{\rm (d)}_{np}(v_p,v_n)
 \,R^{(np)}_p(v_p)\, / \, R^{\rm (d)}_p(v_p)$.

Apart from the neutrino luminosity of the NS core, we take
into account the
neutrino luminosity of the crust resulting
from the bremsstrahlung neutrino emission
of electrons scattered by atomic nuclei.
In this case, the approximate formula proposed by Maxwell (1979)
is used.

The heat capacity of the NS is
assumed to be the sum of the $n, \, p$, and $e$ heat
capacities in the stellar core; we neglect
the heat capacity of the crust because
the crust mass is low in the chosen NS models. The effects of
superfluidity on the $n$ and $p$ heat capacities were
investigated by Levenfish
and Yakovlev (1993). When the
superfluidity appears, the heat capacity experiences a
jump due to the appearance of the
latent heat. The proton (singlet superfluidity)
and the neutron (triplet superfluidity)
jump abruptly by factors of 2.43 and 2.19,
respectively. At $T \ll T_c$, the nucleon heat capacity is exponentially
suppressed. In our calculations of the
neutrino luminosity and heat capacity, we
assume
the effective masses of neutrons and protons in the NS core to be 0.7 of
the masses of bare particles.

\section{RESULTS}

We calculated about 1800 cooling curves
for two NS models (Section~2): with the
masses $M = 1.44\, M_\odot$ (enhanced cooling)
and $M = 1.3\, M_\odot$ (standard cooling).
Each curve gives the temporal dependence of the
effective surface temperature $T_s =\te \sqrt{1 - {\cal R}_g/{\cal R}}$
of the NS as detected
by a distant observer, ${\cal R}_g$ being the gravitational
radius of the NS. The curves were calculated
in a wide range of critical
temperatures of neutrons ($\tn$) and protons (\tp) in the NS core
(from $10^7$ to $10^{10}$~K).

Typical cooling curves are plotted in Fig.~1.
Curve 2 corresponds to the standard
neutrino losses ($M = 1.3\, M_\odot$) in a NS with the
normal (nonsuperfluid) core. The
change in the shape of the curve at
$t = t_\nu  \approx 2.5 \times 10^5$~yr reflects the
change of the cooling regime.
For $t \la t_\nu$, the NS is at the stage of
{\it neutrino} cooling: the neutrino luminosity of the inner
stellar layers is
considerably higher than the photon luminosity
of its surface. For $t \ga t_\nu$,
the neutrino luminosity falls well below the photon one,
and the {\it photon} stage sets in. Owing to the low neutrino luminosity,
the cooling proceeds rather slowly.
Curve 1 corresponds to the
enhanced neutrino losses ($M = 1.44 \, M_\odot$) in the
NS with a normal core. In this case, the NS
cools much faster, with the neutrino-cooling stage being slightly
delayed. The duration $t_\nu$ of the
neutrino stage is mainly determined
by the stellar heat capacity and by the
dependence of the neutrino luminosity on the NS core temperature.
For normal nucleons, $t_\nu \sim (2$--$ 4) \times 10^5$~yr.
A strong proton superfluidity with normal neutrons
reduces the heat capacity and time $t_\nu$ by about (20--30)\%.
For a strong neutron superfluidity and normal protons,
the heat capacity and $t_\nu$ decrease
by a factor of 2 to 3. In the case of strong $n$ and $p$ superfluidities,
the heat capacity drops approximately
by a factor of 20, and $t_\nu \sim 10^4$~yr.

\begin{figure*}[h]
\begin{center}
\leavevmode
\epsfysize=120mm
\epsfbox[67 230 532 570]{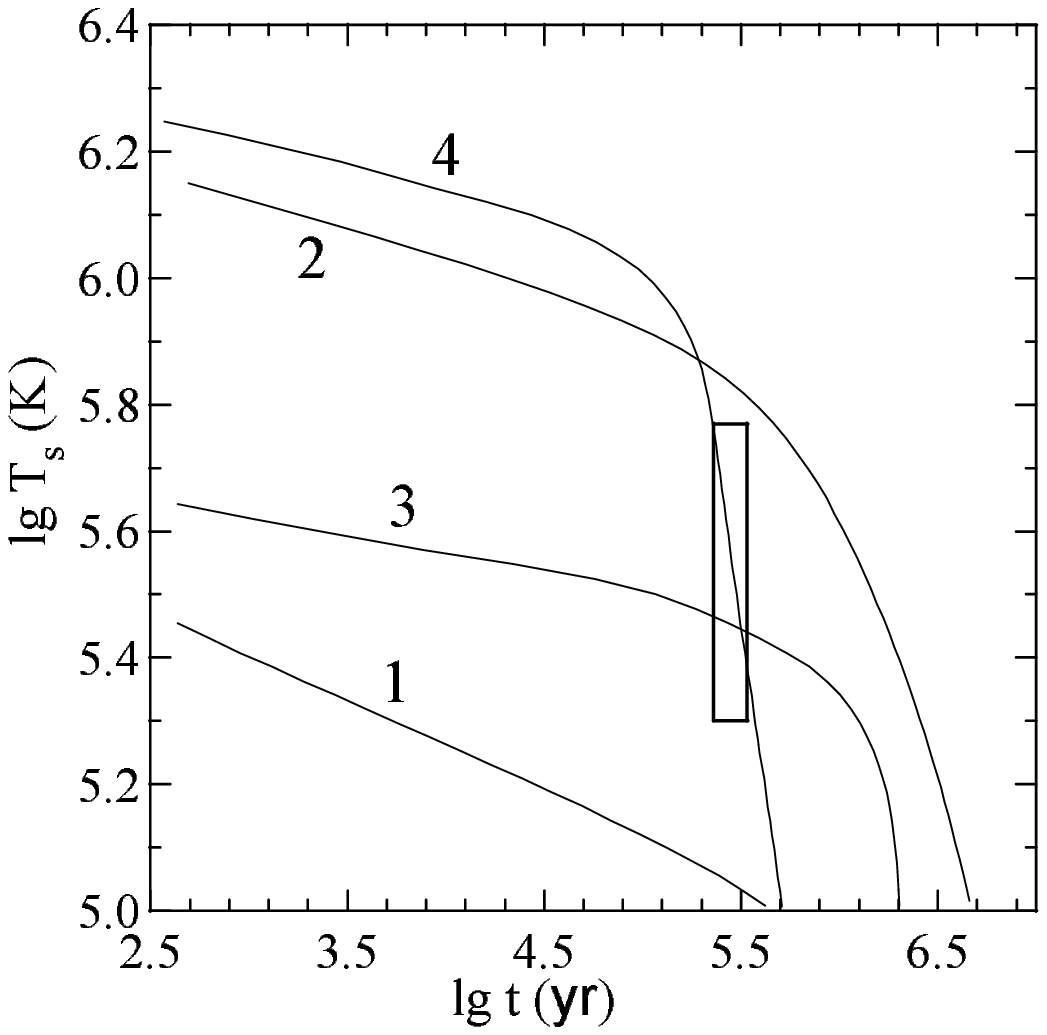}
\end{center}
\caption[]{\label{}
Cooling curves for a neutron star with enhanced (curves 1 and 3)
and standard (curves 2 and 4)
neutrino luminosities. Curves 1 and 2 correspond to
the absence of superfluidity; curve 4
to $\tn = 10^{9.5}$~K, $\tp = 10^{8.5}$~K; and curve 3
to $\tn = 10^{7.3}$~K, $\tp = 10^{8.5}$~K. The rectangle encloses the region
of the selected effective surface temperatures $\te$ and Geminga's ages $t$.
}
\nonumber
\end{figure*}

The results of our calculations are
of general character. We will analyze them
using the Geminga pulsar as an example.
Recent observations of Geminga from the
ROSAT X-ray  Orbital Observatory provide
a wealth of information on the X-ray
emission of this pulsar (Halpern and Holt 1992;
Halpern and Ruderman 1993; Meyer
{\it et al.} 1994). The emission consists of soft and hard components. The soft
component is interpreted as the thermal emission from the Geminga
surface, while the hard component is commonly attributed to the emission of
small hot spots on the stellar surface. Theoretical interpretation of the
observational data is ambiguous.
Thus, for example, Halpern and Ruderman (1993)
described the soft and hard components
by blackbody spectra with different
temperatures and estimated the effective
surface temperature to be $\te = (5.2 \pm 1.0) \times 10^5$~K.
The same authors fitted the soft and hard components with a
blackbody spectrum and a power-law spectrum, respectively.
In this model,
$\te = (4.5 \pm 1.4) \times 10^5$~K. Meyer {\it et al.} (1994)
interpreted the emission of the hard
component using a set of NS hydrogen-atmosphere models
with a magnetic field.
The best-fit models are consistent with a lower temperature,
$\te = (2$--$3) \times 10^5$~K. Recently, Page {\it et al.} (1995) have
interpreted the spectroscopic observations of Geminga
using the model of a cool hydrogen
nonmagnetic atmosphere ($\te \approx 2.2 \times 10^5$~K) without
a magnetic field
in the presence of two small hot spots ($4 \times
10^5$~K) with a radial magnetic field of opposite polarity.
It should be stressed that the modeling of NS atmospheres is far
from being complete. It is necessary
to solve a complicated problem of radiative transfer
in a magnetized, sufficiently
cool and dense (nonideal) plasma of
the NS atmosphere with a complex chemical
composition (hydrogen, helium, and heavy elements).
It is quite possible that
new models and observational data will
yield slightly different values of $\te$.
We will not favor any of the available interpretations,
but consider the temperature $\te$
as an unknown quantity. We will take $\te = (2$--$6) \times 10^5$~K
as the most plausible temperature $T_s$ (with allowance for the gravitational
redshift). For the $M = 1.3\, M_\odot$ and $M = 1.44\, M_\odot$ models,
the $T_s/\te$ ratios are equal to 0.820 and 0.791, respectively.

The age of Geminga is estimated
(Bertsch {\it et al.\ } 1992; Page 1994) from the
observed rotation period $P \approx 0.237$~s and
its  time derivative $\dot{P} = 1.10 \times
10^{-14}$~s/s using the well-known formula
$t=(P/\dot{P})/(n-1)$, where $n$ is the braking index.
For the magnetic-dipole losses ($n=3$),
$t=t_1=3.4 \times 10^5$~yr. For $n=4$, we have
$t=t_2=2.3 \times 10^5$~yr, which may be considered
as the lower limit of the Geminga's age (Page 1994).

We will assume the ranges of surface
temperatures $T_s = (2$--$ 6) \times 10^5$~K and ages
$t_2 \le t \le t_1$ (Fig.~1) to form an "error box"
which the real cooling curves
of the Geminga pulsar must intersect. One can see that in the absence of
superfluidity, the cooling proceeds
more slowly than required for the standard
neutrino luminosity and more rapidly
than required for the enhanced luminosity.
A detailed analysis made by Page (1994)
has shown that both the standard and enhanced neutrino luminosities
fit the observations if the effect of
superfluidity is taken into account.
This assertion is illustrated by curves~3
and 4 in Fig.~1. Curve~3 corresponds to
the standard neutrino energy losses ($M = 1.3\, M_\odot$) at
$\tn = 10^{9.5}$~K and $\tp = 10^{8.5}$~K.
Such a superfluidity decreases  strongly
the heat capacity of the star and accelerates the
cooling at the photon stage (when
the superfluid suppression of the neutrino luminosity is
insignificant). Curve~4 corresponds to the
enhanced neutrino luminosity ($M =
1.44 \, M_\odot$) at $\tn = 10^{7.3}$~K and $\tp = 10^{8.5}$~K.
The superfluidity chosen
suppresses the neutrino luminosity more strongly
than the heat capacity, and delays the cooling at the neutrino stage
and at the transition stage from the neutrino to
photon cooling.

With the inclusion of new data on neutrino
energy losses and the NS heat capacity (Sections~1 and 2), our
calculations extend the study of Page (1994) to
a wider range of $\tn$ and $\tp$.
In view of the great variety of the results,
their representation by cooling curves is inconvenient.
It is more convenient
(Gnedin {\it et al.} 1994) to display the critical
superfluidity temperatures $\tn$
and $\tp$ which lead to definite effective
surface temperatures $T_s$ of a NS of given
age $t$ (Figs.~2--5). Figures~2 and 4 correspond to the age
$t_1$, while Figs.~3
and 5 correspond to the age $t_2$.

Figures~2 and 3 correspond to enhanced neutrino energy losses.
Since the figures
are similar, we will describe Fig.~2 ($t = t_1$)
as an example.  The lower left
corner of this figure corresponds to the case where the critical temperatures
$\tn$ and $\tp$ are below or of the same order of magnitude as the
internal temperature of a NS of age $t_1$. The nucleon superfluidity
either has not yet appeared
in or has had no time to strongly affect the cooling. As a result, the NS
surface temperature drops to $ \sim 10^5 $~K (Fig.~1)
through intense neutrino energy
losses in the direct Urca process, with the star being still at the
neutrino cooling stage
($t_\nu \ga t_1$).
\begin{figure}
\begin{center}
\leavevmode
\epsfysize=100mm
\epsfbox[60 120 530 570]{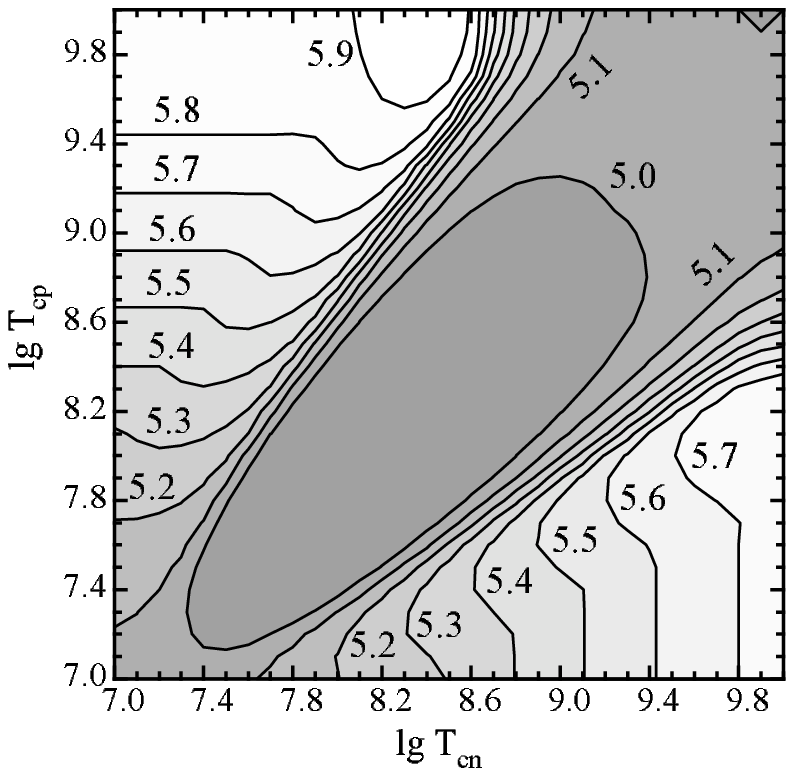}
\end{center}
\caption[]{\label{figure2}
$\tn$ and $\tp$ lines corresponding to
certain surface temperatures $T_s$ of
a neutron star with the
enhanced neutrino luminosity ($M = 1.44\, M_\odot$) and age of
$3.4 \times 10^5$~yr. The numbers near
the curves are the values of $\log T_s$.
}
\end{figure}

In all the remaining parts of Fig.~2,  the effect of superfluidity on the
cooling turns out to be crucial.
Calculations and a simplified analytic analysis
of the cooling equations show
that $t_\nu \sim t_1$ throughout the figure (with the
exception of the upper right corner of Fig.~2). In that case, the cooling is
governed by both the neutrino luminosity and the heat capacity of the NS.
The neutron superfluidity affects
the direct Urca process and the proton heat capacity
in about the same manner
as the proton superfluidity affects the direct
Urca process and the proton heat capacity
(Levenfish and Yakovlev 1994a,b). In
addition, the heat capacities of normal neutrons
and protons are comparable but
considerably larger
than the electron heat capacity. As a result, the curves in Fig.~2
are almost
symmetric with respect to the replacement of the coordinate axes,
$\log \tn \leftrightarrow \log \tp$, and
it is sufficient, for instance, to describe the  lower right half
of Fig.~2.

If a low temperature of $\tp \sim 10^7$~K is fixed
and $\tn$ increases from $\sim 10^8$
to $\sim 10^{10}$~K, the proton superfluidity does not
occur by the time $t_1$, and the
cooling is governed by the neutron superfluidity alone. Therefore, the
$T_s = {\rm const}$ lines in the lower right corner of Fig.~2
are vertical. As $\tn$ increases,
the neutrino luminosity drops by several
orders of magnitude. The heat capacity
also changes but much weaker,
because the strong $n$ superfluidity
reduces the NS heat capacity only by a factor of approximately 3.
As a result,
variations in the neutrino luminosity
are more pronounced than variations in the
heat capacity. The cooling of the star
slows down, while $T_s$ rises. If, on the
other hand, we fix $\tn \ga 10^8$~K and increase
$\tp$, then the $p$ superfluidity will arise in a star of age $t_1$ at
some value of $\tp$ (and the $T_s = {\rm const}$ line
will no longer be straight).
As $\tp$ increases further, the superfluidity
suppresses the neutrino luminosity and influences the heat capacity. However,
the additional suppression of the direct Urca process by a moderate proton
superfluidity in the presence of a strong neutron
superfluidity is weaker than
the change in the proton heat capacity (Levenfish and Yakovlev 1994a,b).
Then the cooling is mainly controlled by the proton heat capacity.
When the superfluidity arises, the heat capacity
increases in a jump-like
manner
(the latent heat is liberated), giving rise to an increase in $T_s$:
the $T_s = {\rm const}$ lines are shifted leftward.
However,
the increase in the heat capacity is
superseded by its sharp suppression with rising $\tp$.
The NS becomes cooler, and
the isolines are shifted rightward.

Finally, the critical temperatures $\tn$ and $\tp$ in the upper right corner of
Fig.~2 are so large
that the neutron and proton superfluidities suppress the
direct Urca process and the nucleon heat capacity almost
completely. For a very
low (electron) heat capacity, the neutrino-cooling
stage is short ($t_\nu \sim 10^4$~yr, see above). As a result, the NS
had long cooled down by the time $t_1$ through the
photon emission, and its temperature
is very low, $T_s \sim 10^5$~K.

\begin{figure}[h]
\begin{center}
\leavevmode
\epsfxsize=100mm
\epsfbox[60 120 500 550]{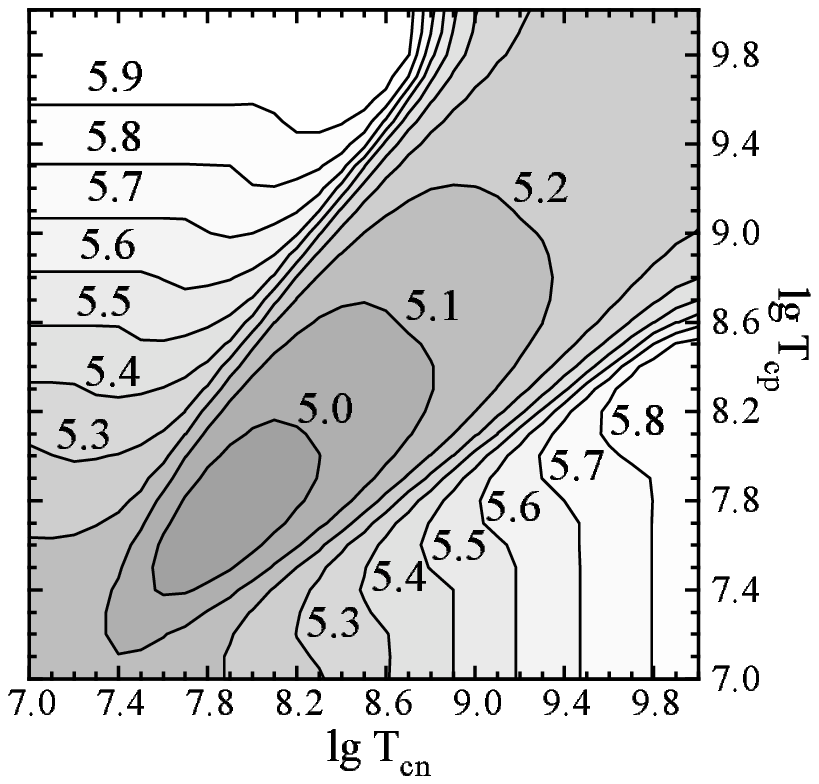}
\end{center}
\caption{\label{figure3}
Same as in Fig.~2 but for a star of age $t = 2.3 \times 10^5$~years.
}
\end{figure}

Figures~2 and 3, corresponding to the enhanced cooling of Geminga, are
similar to Figs.~ 2,~ 4, and 6 from Gnedin {\it et al.} (1994)
composed for the younger
pulsar PSR$\, 0656+14$ ($t \approx 10^5$~yr). A comparison of these
figures shows that
higher critical temperatures of one of the nucleon components under the
condition $\tp \ll \tn$ or $\tn \ll \tp$ are required for
maintaining the high effective temperature $T_s$
over a longer period of time.

$T_s$ isolines for the NS model
with $M = 1.3 \, M_\odot$ are plotted in Figs.~4 and
5. Figures~4 and 5 correspond to the ages $t_1$ and $t_2$,
respectively. It can be
seen that the behavior of the isolines for the standard neutrino energy
losses differs qualitatively from that
for the enhanced neutrino energy losses (Figs.~2
and 3). The principal difference lies in the fact that the relatively weak
standard neutrino luminosity of a NS of age $t_1$ or $t_2$ turns out to be
lower than the photon
luminosity ($t_\nu \la t_2$). The NS of the above age is at the
photon-cooling stage virtually for all values of $\tn$ and $\tp$ shown in Figs.~4
and 5, and the neutrino luminosity of the stellar core plays no significant
role. The nucleon superfluidity affects the cooling mainly through the heat
capacity. The exception is only
some small parts of Figs.~4 and 5.

\begin{figure}[ht]
\begin{center}
\leavevmode
\epsfxsize=100mm
\epsfbox[60 120 500 540]{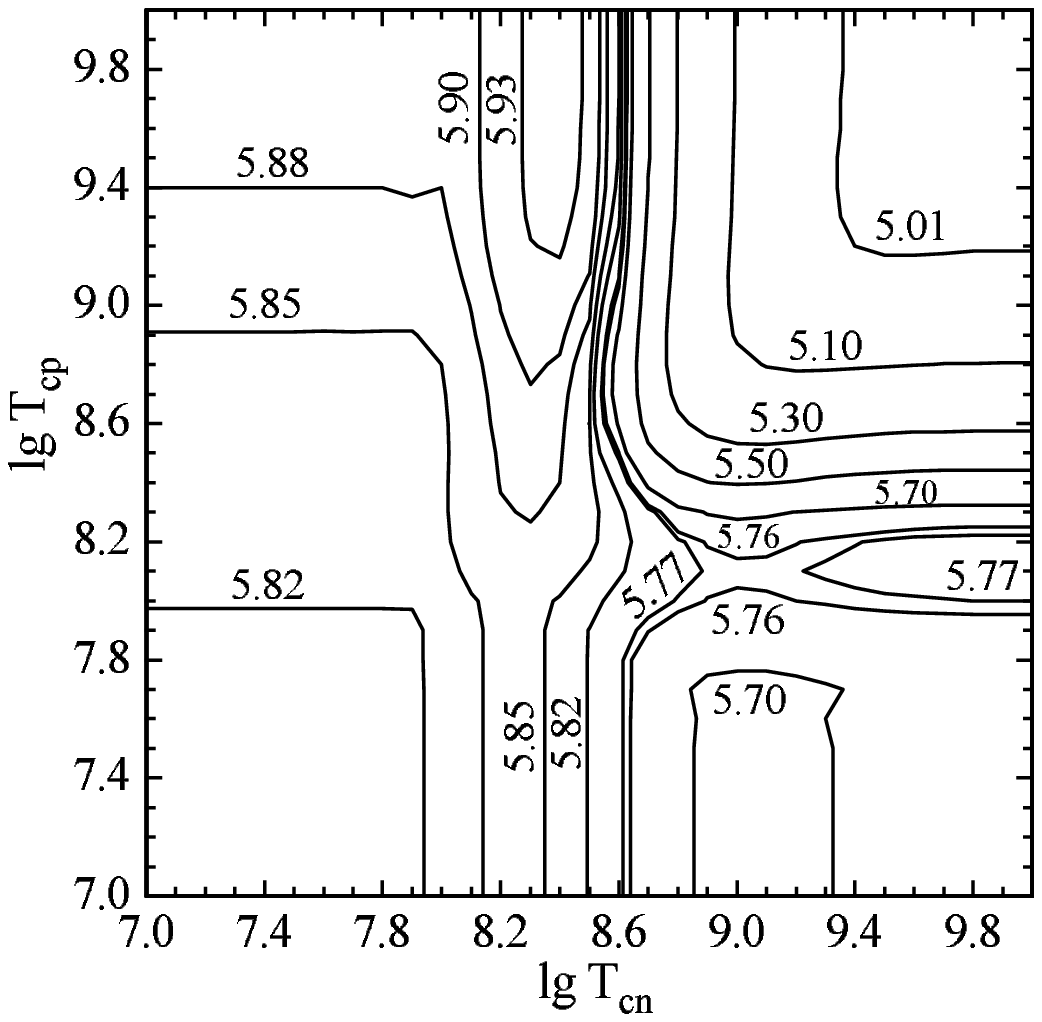}
\end{center}
\caption{\label{figure4}
Same as in Fig.~2 but for a star with standard neutrino luminosity.
}
\end{figure}

\begin{figure}[ht]
\begin{center}
\leavevmode
\epsfxsize=100mm
\epsfbox[60 120 320 370]{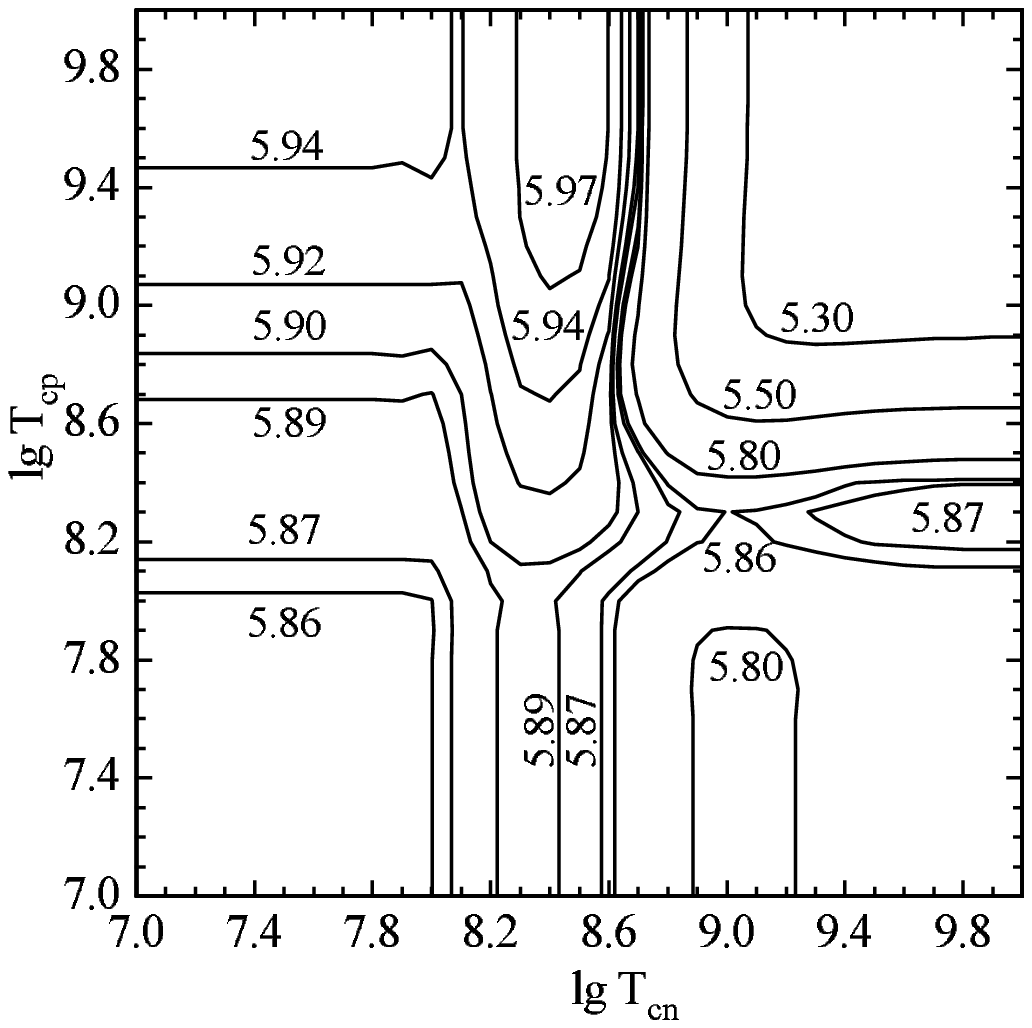}
\end{center}
\caption{\label{figure5}
Same as in Fig.~3 but for the star with enhanced neutrino luminosity.
}
\end{figure}

Figures~4 and 5 are qualitatively similar. We will describe Fig.~4 as an
example. At $\tn < 10^{7.9}$~K and $\tp < 10^{7.9}$~K
(lower left corner of Fig.~4),
the nucleon superfluidity in the core of a NS of age $t_1$
has not yet appeared, and
the surface temperature is independent of $\tn$ and $\tp$. At $\tp
\la 10^{7.5}$~K
(lower part of Fig.~4), the protons are normal.
If, at the same time, we raise
$\tn$, the neutron superfluidity will arise at $ \tn > 10^{7.9}$~K
by the time $t_1$.
At $\tn \la 10^8$~K, this superfluidity is moderate. The latent heat
is released, and $T_s$ slightly rises. As $\tn$ increases
further, the $n$ superfluidity becomes
strong by the time $t_1$; it sharply suppresses
the neutron heat capacity, and decreases
$T_s$. As $\tn$ rises, the $n$ superfluidity arises increasingly earlier.
At $\tn \sim 10^9$~K, it appears at the early neutrino-cooling
stage and is able to strongly
suppress the neutrino losses and
delay the cooling. Though weakly, this delay
manifests itself by the time $t_1$: as $\tn$ becomes higher than $10^9$~K,
the temperature $T_s$ slightly rises.
If, however, we fix $\tn \sim 10^9$~K and raise
$\tp$, the $p$ superfluidity arises at
$\tp \ga 10^{7.6}$~K. The latent heat is
released, and $T_s$ grows. As a result, the
$T_s = {\rm const}$ lines in Fig.~4 form loops
(e.g., $\log T_s = 5.7$).

At $\tn \la 10^{7.9}$~K (left part of Fig.~4),
the neutrons are normal. If $\tp$ is
raised above $\ga 10^{7.9}$~K, the proton superfluidity appears by
the time $t_1$. At
$\tp \sim 10^8$~K, the superfluidity arises just before
$t_1$ and is moderate. The heat capacity increases due
to the latent-heat release, and $T_s$ slightly rises. As
$\tp$ grows further, the $p$ superfluidity
arises at increasingly earlier stages and
becomes strong by the time $t_1$. It suppresses the proton heat capacity,
which amounts to $\la 30 \%$ of the total heat capacity
of the NS for normal neutrons. On
the other hand, this superfluidity exponentially suppresses the neutrino
luminosity at the preceding neutrino-cooling stage.
The slowdown of the cooling
due to the suppression of the neutrino luminosity is more pronounced than the
acceleration caused by an insignificant suppression of the heat capacity, and
$T_s$ rises with increasing $\tp$.

At $\tp \ga 10^9$~K (upper part of Fig.~4), the $p$ heat capacity is
suppressed completely. If $\tn$ is raised, the $n$ superfluidity
appears by the time $t_1$, which first
enhances and then suppresses the $n$ heat capacity (see above).
This causes $T_s$ to initially
rise and then decrease with a peak at $\tn \approx 10^{8.3}$~K.
If $\tn \approx 10^{8.3}$~K is fixed and $\tp$ decreases, $T_s$ slightly drops.
This drop of
$T_s$ is caused by the enhancement of the stellar neutrino luminosity at the
preceding neutrino-cooling stage.
As a result, the $\log T_s = 5.9$ curve forms, for
example, a loop in the upper part of Fig.~4. The $\log T_s = 5.77$ loop
in the right part of Fig.~4 is explained in a similar way
(with replacement $n \leftrightarrow p$).

Finally, at $\tn \ga 10^{8.6}$~K and $\tp \ga 10^{8.3}$~K
(upper right part of Fig.~4),
the $n$ and $p$ superfluidities become strong
by the time $t_1$ and heavily suppress
the nucleon heat capacity. The neutrino-cooling
stage is considerably shorter
($t_\nu \ll t_1$, see above), and the $T_s = {\rm const}$ curves
are entirely determined by a
decrease in the heat capacity
caused by superfluidity. The higher $T_c$, the lower
the heat capacity, and the cooler the star. At $\tn \ga 10^{9.2}$~K, the
$n$ superfluidity suppresses the neutron heat capacity
almost completely, and the
cooling is governed only by a change in $\tp$ (proton heat capacity) at
$\tp  \la 10^9$~K. If, however, $\tp \ga 10^9$~K, the
$p$ superfluidity suppresses the proton
heat capacity almost completely,
and the cooling is governed by a change in $\tn$
(neutron heat capacity) at
$\tn \la 10^{9.2}$~K. If the combined superfluidity of
neutrons and protons is very strong
($\tn > 10^{9.2}$~K, $\tp > 10^{9.1}$~K, see the
uppermost right corner of Fig.~4), the nucleon heat capacity
of the NS core is
fully suppressed, and the cooling no longer depends on $\tn$ and $\tp$. In this
regime, the cooling is governed by the electron heat capacity, which is
independent of the nucleon superfluidity.

It should be emphasized that the
$T_s = {\rm const}$ curves at $\tn \ga 10^{8.5}$~K and $\tp
\ga 10^{8.3}$~K (upper right part of Fig.~4) are
determined by an abrupt suppression
of the heat capacity by strong superfluidity and
are only slightly sensitive to
the specific parameters of the NS. In the remaining region of $\tn$ and $\tp$,
variations in $T_s$ are small, but the $T_s = {\rm const}$
curves (for instance, loops)
may formally strongly deform with changing NS parameters
(cf., the $\log T_s = 5.7$
and $\log T_s = 5.8$ isolines in Figs.~4 and 5).
This deformation is not accompanied
by large variations in $T_s$ and does not affect the main
conclusions concerning
the effect of superfluidity on the cooling.

Note that Page (1994) considered a wider age
range for Geminga, $t_2 \le t \le t_3$,
where $t_3 = 6.8 \times 10^5$~yr corresponds to
the braking index $n = 2$ (see above).
We give no results for $t = t_3$.
In the case  of the
enhanced neutrino losses, the
corresponding figures are similar to Figs.~2 and 3. In the case
of the standard
energy losses, at $t = t_3$, Geminga is at
the advanced photon-cooling stage, when
the surface temperature is sufficiently low
($T_s \approx 5.7 \times 10^5$~K) even for normal
nucleons (Fig.~1).

\section{CONCLUSIONS}
Our calculations support the conclusion of
Page (1994):  observational
data on the thermal emission of Geminga
can be explained by both the standard or
enhanced neutrino energy losses if neutrons
and protons in the
Geminga's core are
assumed to be superfluid.
Figures~2 and 3 give the critical temperatures of the
neutron and proton superfluidities, $\tn$ and $\tp$, which lead to
specified effective surface temperatures $T_s$ of Geminga
if its age is $t_1 = 3.4 \times 10^5$~yr
or $t_2 = 2.3 \times 10^5$~yr (Section~3)
and the neutrino luminosity is enhanced by
the direct Urca process (\ref{eq:Durca}).
Figures~4 and 5 are similar but correspond to the
assumption that the neutrino luminosity is standard
and results from the reactions
(\ref{eq:Murca}) and (\ref{eq:Brema}). Observational data on
the X-ray emission of Geminga (Section~3)
indicate that its effective surface temperature $T_s$
ranges from $2 \times 10^5$~K to $6 \times 10^5$~K.

Both assumptions about the neutrino energy
losses (standard and enhanced) can
be reconciled with observational data.
In both cases, the required values of $\tp$ and $\tn$ are
in qualitative
agreement with various model
calculations of the critical temperatures
of nucleon superfluidity in NS cores
(see Page and Applegate 1992, for references). However, the properties of
superfluidity in the
Geminga's core for the standard and enhanced neutrino
luminosities must greatly differ.
In the case of the enhanced neutrino luminosity,
the superfluidity is needed to slow down
the NS cooling at the neutrino-cooling stage (or at
the transition stage from the neutrino to photon luminosity)
through a partial suppression of the
neutrino energy losses in the direct Urca
process. For example, the values of $\tn$ and $\tp$
that give the Geminga's surface
temperature $T_s = 10^{5.3}$--$ 10^{5.5}$~K lie
(see Figs.~2 and 3) in two relatively
wide decoupled regions in the ($\tn$--$\tp$) plane. For the standard neutrino
processes, the superfluidity must slow down the cooling at the photon-cooling
stage through the suppression of the stellar heat capacity. The region of the
necessary values of $\tn$ and $\tp$ (Figs.~4 and 5)
turns out to be much narrower
and coupled. For the same $T_s$, we obtain
$\log \tn \ga 8.7$ and $\log \tp \ga 8.5$.
In this case, the constraints on possible values of critical temperatures
are much more stringent.
On the whole, the model
of the standard neutrino losses requires higher $\tn$ and $\tp$ than the model
of the enhanced losses.

The results of our calculations are tentative.
More detailed models for the cooling of NSs with superfluid cores are needed.
First, it would
be desirable to allow for the nonuniform distribution of the critical
temperatures $\tn$ and $\tp$ (for their density
dependence) throughout the stellar core. For this to be done,
joint theoretical calculations of $\tn$ and $\tp$ are
necessary for various models of superdense matter
(which are now not available).
These data could be used to directly determine the best-fit models without
assuming that $\tn$ and $\tp$ are free parameters. It would be desirable to
consider the cases where other particles, above all, muons and hyperons, are
present in the NS cores along with neutrons,
protons and electrons.
Note that
the neutrino-loss rates in various types of reaction are also sensitive to
superdense-matter models and have not yet been determined with confidence. At
the late cooling stage, additional heating sources may also be important, for
example, those related to dissipation of
the rotational energy in the core of a
star during its slowdown (see, e.g., Cheng {\it et al.} 1992, Umeda {\it et al.}
1994, and references therein). One may hope
that new observational data on the
thermal emission of NSs together with
new theoretical calculations will make it
possible to obtain more stringent constraints
on the rate of neutrino energy
losses and the critical temperatures of
nucleon superfluidity in superdense matter. \\[2ex]

\noindent
{\it ACKNOWLEDGMENTS}

We wish to thank O.Yu.\ Gnedin for useful
discussions and great  assistance in
our calculations, and to V.S.Imshennik and
Yu.A.\ Shibanov for valuable critical remarks. This work
was supported in part by the
Russian Basic Research Foundation
(project code 93-02-2916), the Soros International
Science Foundation (Grant
R6A-000), the ESO C\&EE Programme
(Grant A-01-068) and by INTAS (Grant 94-3834). \\[2ex]

\noindent
{\it REFERENCES}

\noindent
 Anderson, S.B., C\'{o}rdova, F.A., Pavlov, G.G., {\it et al.},
{\it Astrophys. J.}, 1993, v. 414, p. 867.

\noindent
 Bertsch, D.L., Brazier, K.T.S., Fichtel, C.E.,
{\it et al.}, {\it Nature}, 1992, v. 357, p. 306.

\noindent
 Cheng, K.S., Chau, W.Y., Zhang, J.L., and Chau, H.F.,
{\it Astrophys. J.}, 1992, v. 396, p. 135.

\noindent
 Finley, J.P., \"{O}gelman, H., and Kizilogl\u{u}, \"{U}.,
{\it Astrophys. J.}, 1992, v. 394, p. L21.

\noindent
 Friman, B.L. and Maxwell, O.V.,
 {\it Astrophys. J.}, 1979, v. 232, p. 541.

\noindent
 Glen, G. and Sutherland, P., {\it Astrophys. J.}, 1980, v. 239, p. 671.

\noindent
 Gnedin, O.Yu. and Yakovlev, D.G.,
{\it Astron. Lett.}, 1993, v. 19, p. 104.

\noindent
 Gnedin, O.Yu., Yakovlev, D.G., and Shibanov, Yu.A.,
{\it Astron. Lett.}, 1994, v. 20, p. 409.

\noindent
 Halpern, J.P. and Ruderman, M., {\it Astrophys. J.}, 1993, v. 415, p. 286.

\noindent
 Halpern, J.P. and Holt, S.S., {\it Nature}, 1992, v. 357, p. 222.

\noindent
 Lattimer, J.M., Pethick, C.J., Prakash, M., and Haensel, P.,
{\it Phys. Rev. Lett.}, 1991, v. 66, p. 2701.

\noindent
 Levenfish, K.P. and Yakovlev, D.G.,
Strongly Coupled Plasma Physics, Van Horn,
H. and Ichimaru, S., Eds., Rochester: University Rochester, 1993, p. 167.

\noindent
 Levenfish, K.P. and Yakovlev, D.G.,
{\it Astron. Lett.}, 1994a, v. 20, p. 43.

\noindent
 Levenfish, K.P. and Yakovlev, D.G.,
{\it  Astron. Rep.}, 1994b, v. 38, p. 247.

\noindent
 Maxwell, O.V., {\it Astrophys. J.}, 1979, v. 231, p. 201.

\noindent
 Meyer, R.D., Pavlov, G.G., and M\`{e}sz\'{a}ros, P.,
{\it Astrophys. J.}, 1994, v. 433, p. 265.

\noindent
 Nomoto, K. and Tsuruta, S., {\it Astrophys. J.}, 1987, v. 312, p. 711.

\noindent
 \"{O}gelman, H. and Zimmermann, H.-U.,
{\it Astron. Astrophys.}, 1989, v. 214, p. 179.

\noindent
 Page, D., {\it Astrophys. J.}, 1994, v. 428, p. 250.

\noindent
 Page, D. and Applegate, J.H.,
{\it Astrophys. J. Lett.}, 1992, v. 394, p. L17.

\noindent
 Page, D., Shibanov, Yu.A., and Zavlin, V.E.,
{\it Astrophys. J. Lett.}, 1995, v. 451, L21.

\noindent
 Prakash, M., Ainsworth, T.L., and Lattimer, J.M.,
{\it Phys. Rev. Lett.}, 1988, v. 61, p. 2518.

\noindent
 Shapiro, S.L. and Teukolsky, S.A.,
Black Holes, White Dwarfs, and Neutron Stars, New York: Wiley, 1983.

\noindent
 Shibanov, Yu.A. and Yakovlev, D.G., {\it Astron. Astrophys.}, 1996,
v. 309. p. 171.

\noindent
 Umeda, H., Tsuruta, S., and Nomoto, K.,
{\it Astrophys. J.}, 1994, v. 433, p. 256.

\noindent
 Van Riper, K.A. and Lattimer, J.M.,
in: Isolated Pulsars, Van Riper, K.A. and
Epstein, R., Eds., Cambridge: Cambridge Univ., 1993, p. 122.

\noindent
 Van Riper, K.A., {\it Astrophys. J., Suppl. Ser.}, 1991, v. 75, p. 449.

\noindent
 Wambach, J., Ainsworth, T.L., and Pines, D.,
in: Neutron Stars: Theory and Observation, Ventura, J. and Pines, D., Eds.,
Dordrecht: Kluwer Acad., 1991, p. 37.

\noindent
 Yakovlev, D.G. and Kaminker, A.D.,  in:
Equations of State in Astrophysics,
Chabrier, G. and Schatzman, E., Eds., Cambridge: Cambridge Univ., 1994, p. 214.

\noindent
 Yakovlev, D.G. and Levenfish, K.P.,
{\it Astron. Astrophys.}, 1995, v. 297, p. 717. \\[1ex]

\end{document}